# Study of Target Dependence of Clan Model Parameter in $^{84}Kr_{36}$ - Emulsion Interactions at 1 GeV per nucleon

M K SINGH[1,2,3], A K SOMA[3], V SINGH[*1,3], and R PATHAK[2]

1 Physics Department, Banaras Hindu University, Varanasi – 221005, India

2 Physics Department, Tilak Dhari Postgraduate College, Jaunpur – 222002, India

3 Institutes of Physics, Academia Sinica, Taipei – 11529, Taiwan

* Corresponding author. E-mail: venkaz@yahoo.com

**Abstract.** The article focuses on study of clan model parameters and their target dependence in light of void probability scaling for heavy (Ag and Br) and light (C, N and O) groups of targets present in nuclear emulsion detector using $^{84}Kr_{36}$ at ~1 A GeV. The variation of scaled rapidity–gap (rap-gap) probability with single moment combination has been studied. We found that experimental points are lying approximately on the negative binomial distribution (NBD) curve, indicating a scaling behavior. The average clan multiplicities ($\bar{N}$) for interactions, increases with the pseudo-rapidity interval ($\Delta\eta$) was also observed. The values of $\bar{N}$ for AgBr targets are larger than those for CNO target and also average number of particles per clan ($\bar{n}_c$) increases with increase in pseudo-rapidity interval. We further observed that for a particular target, average number of particles per clan ($\bar{n}_c$) increases with an increase in the size of projectile nucleus.

**Keywords:** $^{84}Kr_{36}$-Emulsion Interaction, Nuclear Emulsion Detector, Clan Model Parameters and their Target Dependence.

## 1. Introduction

The search for final state particles produced in nucleon-nucleus or nucleus-nucleus interactions at high energy is the main objective of relativistic heavy ion experiments [1-6]. The multiplicity variable studies elucidates on number of particles have been produced in an interactions and thus quantify the physics behind multi-particle production process. In case of cascade event, any particle produced from a primary cluster termed as ancestor and all the particles having common ancestor form a clan [7]. The clans have no mutual interactions, whereas the particles emitted due to disintegration of a clan have strong correlations. In the context of clan model, average clan multiplicity $\bar{N}$ and average number of particles per clan $\bar{n}_c$ are of immense physical importance and are referred as clan model parameters [7-11]. The clan model parameter can be calculated by rapidity gap (rap-gap) probability scaling, which is also known as void probability scaling [12]. The rap-gap probability is defined as the probability of events with zero particles in a specific region of phase space [12–14]. In earlier research it was reported that rap-gap probability $P_0(\Delta\eta)$ is a useful parameter to study correlations among secondary particles produced in certain interactions [7-11]. In this paper, we have computed clan model parameters and studied their target dependence in light of void probability scaling in nuclear emulsion using $^{84}Kr_{36}$ at ~ 1 A GeV as projectile.

## 2. Experimental details

Data used in this analysis was collected with a stack of high sensitive NIKFI BR-2 nuclear emulsion pencils of dimensions 9.8 × 9.8 × 0.06 cm$^3$, which was exposed horizontally by $^{84}Kr_{36}$ ion of kinetic energy ~1 GeV per nucleon. The exposure was performed at Gesellschaft fur Schwerionenforschung (GSI) Darmstadt, Germany. Interactions were found by along the track scanning technique using an oil immersion objective of 100X magnification. There are two standard methods used for scanning of the emulsion plates, one is the line scanning and other is volume scanning. In line scanning method, beam tracks were picked up at a distance of 5 mm from the edge of plate and are carefully followed until they either interacted with nuclear emulsion detector nuclei or escaped from any surface of emulsion. In volume scanning, emulsion plates are scanned strip by strip and event information was collected [6, 15, 16].

All charge secondary produced in an interaction are classified in accordance with ionization or normalized grain density ($g^*$), range (L) and velocity ($\beta$) into three categories [16-18].

**2.1. Shower tracks ($N_s$):** These are freshly created charged particles with $g^*$ less than 1.4. These particles have $\beta > 0.7$. They are mostly fast pions with a small mixture of Kaons and released protons from the projectile which have undergone an interaction. For the case of proton, kinetic energy ($E_p$) should be less than 400 MeV.

**2.2. Grey tracks ($N_g$):** Particles with range L > 3 mm and $1.4 < g^* < 6.0$ are defined as greys. They have $\beta$ in the range of $0.3 < \beta < 0.7$. These are generally knocked out protons. (NED's can not detect neutral particle) of targets with kinetic energy in between 30 - 400 MeV, and traces of deuterons, tritons and slow mesons.

**2.3. Black tracks ($N_b$):** Particles having L < 3 mm from interaction vertex and g* > 6.0. This corresponds to β < 0.3 and protons of kinetic energy less than 30 MeV. Most of these are produced due to evaporation of residual target nucleus. The number of heavily (h) ionizing charged particles ($N_h$) are part of the target nucleus is equal to the sum of black and gray fragments ($N_h = N_b + N_g$).

NED's are a composite target detector. It composed of mainly H, C, N, O, Ag and Br nuclei. The incident projectile will interact with either one of the targets. These emulsion targets are generally classified into three major classes which are combination of AgBr nuclei having averaged $A_T$ = 94 for heavy; CNO nuclei having averaged $A_T$ = 14 for medium and the free hydrogen nucleus having $A_T$ = 1 for light targets [6, 16]. The number of heavily ionizing charged particles depends upon the target breakup. The target separation was achieved by applying restrictions on the number of heavily ionizing charge particles and on residual range of black particles emitted in each event [6, 16].

**2.4. AgBr target events:** $N_h \geq 8$ and at least one track with R < 10 μm is present in an event.

**2.5. CNO target events:** $2 \leq N_h \leq 8$ and no tracks with R < 10 μm are present in an event. This class always contains very clean interaction of CNO target.

**2.6. H target events:** $N_h \leq 1$ and no tracks with R<10 μm are present in an event. This class include all $^{84}$Kr+H interactions but also some of the peripheral interactions with CNO and the very peripheral interactions with AgBr targets.

3. **The model**

The probability of producing or detecting n number of particles in a pseudorapidity interval $\Delta\eta$ be $P_n(\Delta\eta)$. The probability generating function $Q(\lambda)$ for $P_n(\Delta\eta)$ can be defined as [19],

$$Q(\lambda) = \sum_{n=0}^{\infty} (1-\lambda)^n P_n(\Delta\eta) \quad (1)$$

Where, $\lambda$ is a real variable and is restricted to a suitable convergence domain. $Q(\lambda)$ can be written in terms of the reduced factorial cumulant $\overline{K}_N$ as

$$Q(\lambda) = \exp\left(\sum_{N=1}^{\infty} \frac{(-\lambda\bar{n})^N}{N!} \overline{K}_N\right) \quad (2)$$

Where, $\bar{n}$ is the average number of particles in the $\Delta\eta$ region. Inverting equation (1) we have

$$P_n(\Delta\eta) = \frac{(-1)^n}{n!}\left(\frac{\partial^n Q(\lambda)}{\partial \lambda^n}\right)_{\lambda=1} \quad (3)$$

In order to find the probability of producing zero particles in an interval $\Delta\eta$, we have to substitute n = 0 in equation (3)

$$P_0(\Delta\eta) = Q(\lambda)|_{\lambda=1} \quad (4)$$

Equation (4) depicts relation between $P_0(\Delta\eta)$ and the generating function $Q(\lambda)$. This probability, $P_0(\Delta\eta)$, in turn may be used as the generating function for $P_n$

$$P_n(\Delta\eta) = \frac{(-\bar{n})^n}{n!}\left(\frac{\partial}{\partial\bar{n}}\right)^n P_0(\Delta\eta)$$
(5)

This equation expresses relation between *n*-particle and zero-particle probabilities in a region Δη. Equation (5) has been obtained by allowing only $\bar{n}$ to vary in $P_0(\Delta\eta)$ and all other parameters of $P_0(\Delta\eta)$ are taken to be fixed with respect to variation of $\bar{n}$. The gap probability also relates probability $P_n(\Delta\eta)$ with n ≠ 0 through various kinds of moments. $P_0(\Delta\eta)$ can be written as an expansion in cumulants as

$$\ln P_0(\Delta\eta) = \sum_{N=1}^{\infty} \frac{(-\bar{n})^N}{N!} \overline{K}_N$$
(6)

Applying linked pair ansatz to normalized cumulant moment $K_N$ [20], we get

$$\overline{K}_N = A_N \overline{K}_2^{N-1}$$
(7)

If linking coefficients $A_N$ are independent of collision energy and pseudo-rapidity interval and from confirmations of UA1 and UA5 data up to N = 5 [20] a quantity called the scaled rap-gap probability (χ) can be constructed, such that

$$\chi = \frac{-\ln P_0(\Delta\eta)}{\bar{n}}$$
(8)

Then χ depends only on the product of $\bar{n}$ and $\overline{K}_2$. This entitles one to write

$$\chi = \sum_{N=1}^{\infty} \frac{1}{N!} A_N (-\bar{n}\overline{K}_2)^{N-1} = \chi(-\bar{n}\overline{K}_2)$$
(9)

Equation (8) shows that for any correlation among particles, the function $\chi$ is less than unity. $\chi <1$ is a direct manifestation of clustering of particles. This feature makes scaled gap probability suitable to investigate production of charged pions and their structure in rapidity space.

In clan mode, hadron production in an interaction results from two-step process. In first step, hadronic clusters each of average size $\bar{n}_c$ are emitted independently from source, according to a Poisson distribution with average clan multiplicity $\bar{N}$. This is followed in second step by fragmentation of clans into final-state hadrons yielding $\bar{n} = \bar{N}\bar{n}_c$. Since, clan production is Poissonian, the average clan multiplicity $\bar{N}$ in an interval $\Delta\eta$ is

$$\bar{N} = -\ln P_0(\Delta\eta). \tag{10}$$

The average number of particles per clan is then given by

$$\bar{n}_c = \frac{\bar{n}}{\bar{N}} = \frac{1}{\chi}. \tag{11}$$

The relations (10) and (11) will only be valid, when experimentally measured values of $\chi$ can be fitted with NB distribution. According to this distribution, linking coefficients increase as $A_N = (N-1)!$ and scaled rap-gap probability $\chi$ should satisfy the relation $\chi = \ln(1 + \bar{n}\bar{K}_2)/\bar{n}\bar{K}_2$. This model is called NB model. If the experimental points of variation of $\chi$ against $\bar{n}\bar{K}_2$ can be fitted with NBD model, $\chi = \ln(1+ \bar{n}\bar{K}_2)/\bar{n}\bar{K}_2$, a scaling behavior will be observed. In light of this scaling behavior, we can determine average clan multiplicity $\bar{N}$ and average number of particles per clan $\bar{n}_c$ in an interval $\Delta\eta$ quite easily.

## 4. Results and discussions

The rap-gap probability $P_0(\Delta\eta)$ is calculated for first pseudo-rapidity interval ($\Delta\eta = 1$) centered on zero using $^{84}Kr_{36}$ at ~1 A GeV projectiles colliding with heavy (AgBr), light (CNO) groups of targets present in NED. The value of pseudo-rapidity interval $\Delta\eta$ was then increased in steps of 1 for each interaction and value of $P_0(\Delta\eta)$ is computed. The values of scaled gap probability for each set of target and projectile were calculated by equation (8). To calculate single-moment combination ($\bar{n}\bar{K}_2$), we determined the value of $\bar{K}_2$. $\bar{K}_2$ is given by $\bar{K}_2 = (<F_2> - 1)$, where $<F_2>$ is second order factorial moment and given as $<F_2> = <n(n-1)>$ and n is the number of particles in $\Delta\eta$ region.

According to two-source model, it is assumed that there are two types of source responsible for the multi-particle production one of them is chaotic source described by the NBD model while the second one is coherent source described by minimal model or hierarchical Poisson model [8–11]. If experimental points can be fitted by NBD model then multi-particle production mechanism is chaotic. If experimental points can be fitted with expectations of minimal model then particle production is coherent [7, 8]. However, there is also a possibility that experimental points may lie in region bounded by NBD model and minimal model then particle production mechanism is partly coherent and partly chaotic [7–11, 21].

The variation of $\chi$ with $\bar{n}\bar{K}_2$ is shown in figure 1. The dotted and solid lines represents fitting of NBD model and minimal model, respectively. Figure 1 shows that for all interactions experimental points lie approximately on dotted curve. This reveals that scaling behavior of experimental points is explained by NBD model and thus describes pion multiplicity distribution.

Therefore, multi-particle production mechanism is chaotic. The experimental points for $^{32}$S-AgBr interactions at 200 A GeV; lies in region bounded by NBD model and minimal model fitting curves. This observation suggests that particle production at higher energy is partly chaotic and partly coherent. Hence, particle production mechanism is chaotic for low energy interactions while for higher energy interactions it has partly coherent and partly chaotic. This inference is also reported by other experimental works [7–11, 21–23].

The values of average clan multiplicity in an interval $\Delta\eta$ and average number of particles per clan ($\bar{n}_c$) is calculated from equations (10) and (11). The variation of $\bar{N}$ and $\bar{n}_c$ with $\Delta\eta$ is shown in figures 2 and figures 3, respectively. The values of average clan multiplicity $\bar{N}$, increase with increase in interval $\Delta\eta$ for both kinds of targets is observed from figure 2. The behavior of average clan multiplicity $\bar{N}$ with interval $\Delta\eta$ is similar for two target groups, while the values of $\bar{N}$ for AgBr target are larger than those for CNO target. Such behavior reveals the fact that rap-gap probability values of the AgBr target are considerably less than that of CNO target.

Figure 3, demonstrates that initially there is a systematic increase in average number of particles per clan ($\bar{n}_c$) with $\Delta\eta$ values for both targets and then saturation of $\bar{n}_c$ is observed for low mass projectile ($^{16}$O and $^{22}$Ne) around $\Delta\eta = 6$. While the saturation value of $\bar{n}_c$ increases with increase in projectile size. According to clan model [13, 14] with increase in $\Delta\eta$, more and more D - clans become full clans. The D - clan is a clan which has all or some of its particles in a limited domain D of considered phase space. Hence, $\bar{n}_c$ is expected to grow, leading to a limiting value. Thus, the saturation value of $\bar{n}_c$ is strongly depends on projectile size, which is also reported by other experiments [7–14].

## 5. Conclusions

The study of clan model parameters and their target dependence in terms of void probability scaling has revealed some interesting physics of multi-particle production process with two different kinds of target in NED. Depending on numbers of heavy tracks, total ensembles of events for projectile were divided into heavy (AgBr) and light (CNO) group of targets. The variation of scaled rap-gap probability with single moment combination was studied. The experimental scaled rap-gap probability is found to lie approximately on NBD curve, indicating a scaling behavior. At high energy interactions, experimental point's lies in region bounded by NBD model and minimal model. Therefore, particle production mechanism is chaotic for low energy interactions and is partly coherent and partly chaotic for higher energy interactions.

Average clan multiplicity $\bar{N}$ for all of interactions increases with increase in pseudo-rapidity interval. However, in case of Ag Br target, $\bar{N}$ is higher than those for CNO target. The average number of particles per clan ($\bar{n}_c$) initially increase with increase in pseudo-rapidity interval. The increase in $\bar{n}_c$ with $\Delta\eta$ can be explained in terms of the formation of D - clans. It has also been observed from the present analysis that for a particular target average number of particles per clan ($\bar{n}_c$) increases with an increase in the size of projectile nucleus. Our results are more or less consistent with other experimental works.

## Acknowledgement

The authors are grateful to all the technical staff of GSI, Germany for exposing nuclear emulsion detector with $^{84}Kr_{36}$ beam.

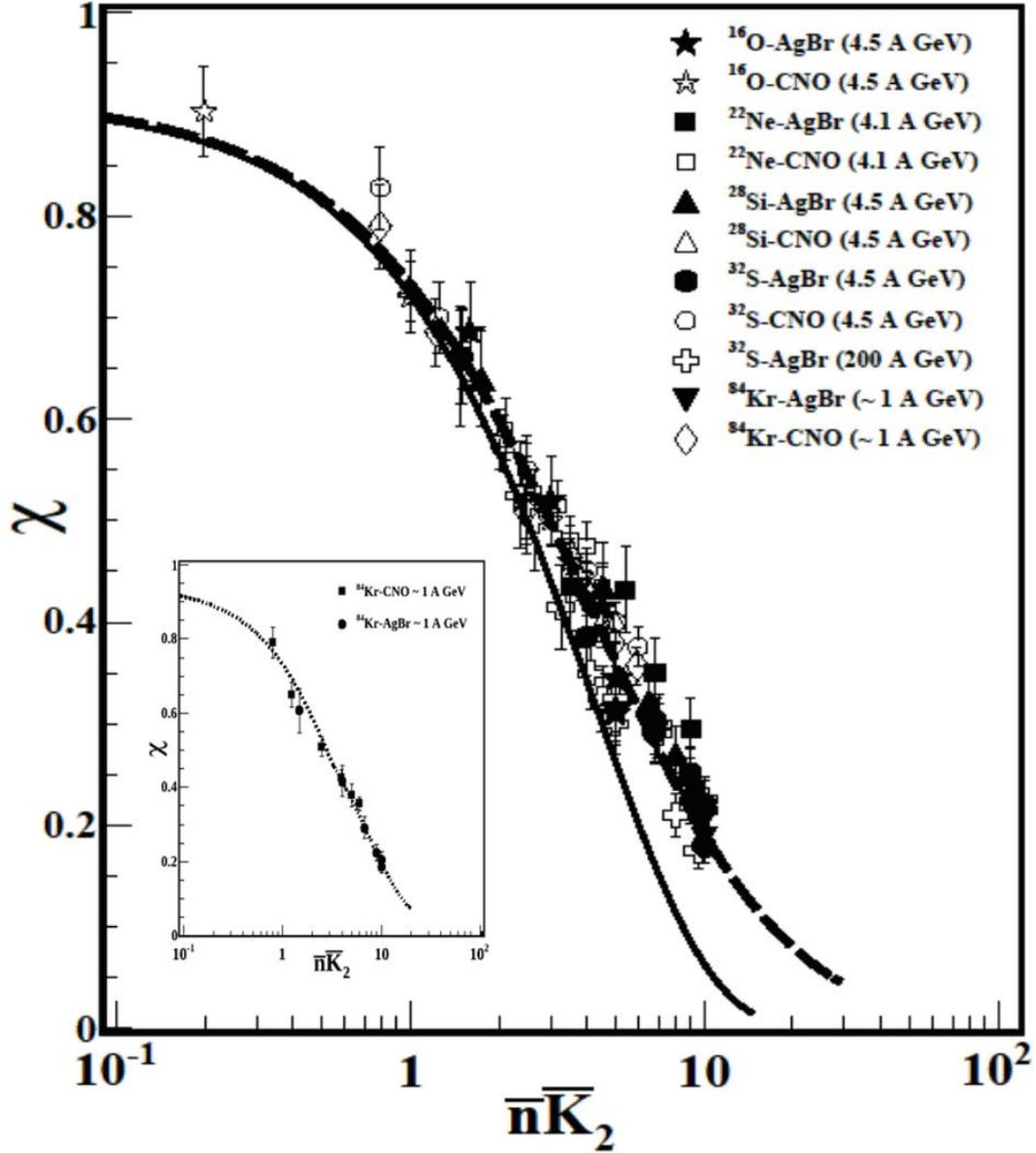

**Figure 1:** The variation of $\chi$ with respect to $\bar{n}\bar{K}_2$. The experimental data points are from $^{16}O$ at 4.5 GeV/c [5], $^{22}Ne$ at 4.1 GeV/c [5], $^{28}Si$ at 4.5 GeV/c [5], $^{32}S$ at 4.5 GeV/c [5], $^{32}S$ at 200 GeV/c [6], $^{84}Kr$ at 0.95 A GeV [**Present work**]. The dotted and solid lines are fitting of data points with NBD model and minimal model, respectively. For clarity, present work is separately shown in the inset.

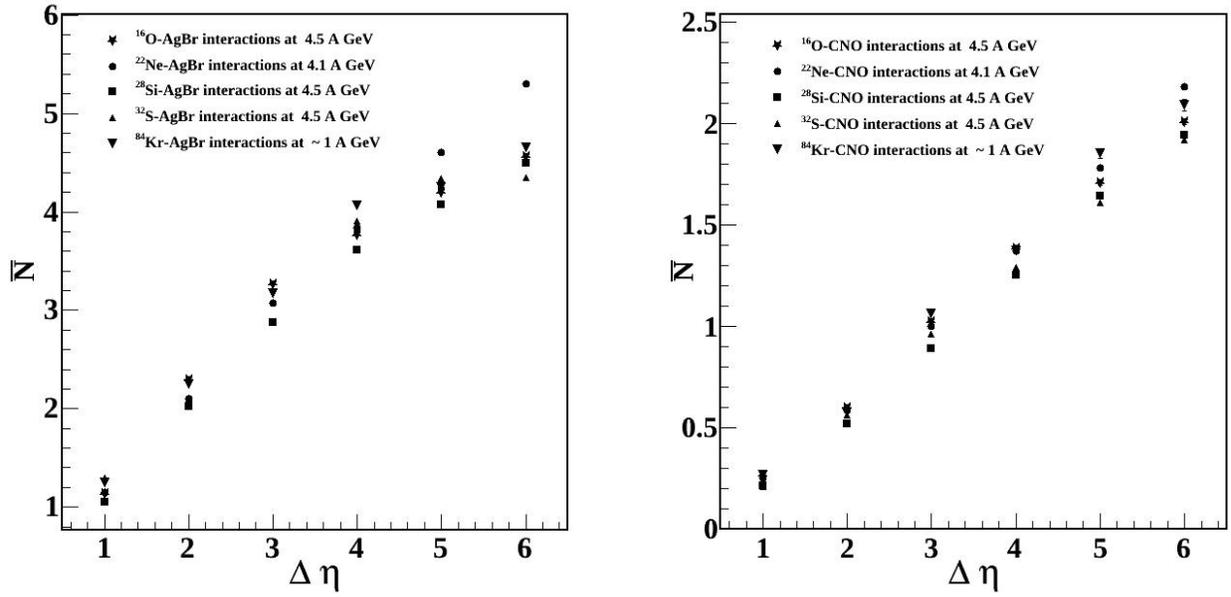

**Figure 2:** The variation of average clan multiplicity ($\bar{N}$) against pseudo-rapidity interval ($\Delta\eta$) for different projectile with Ag Br target (left side) and CNO target (right side).

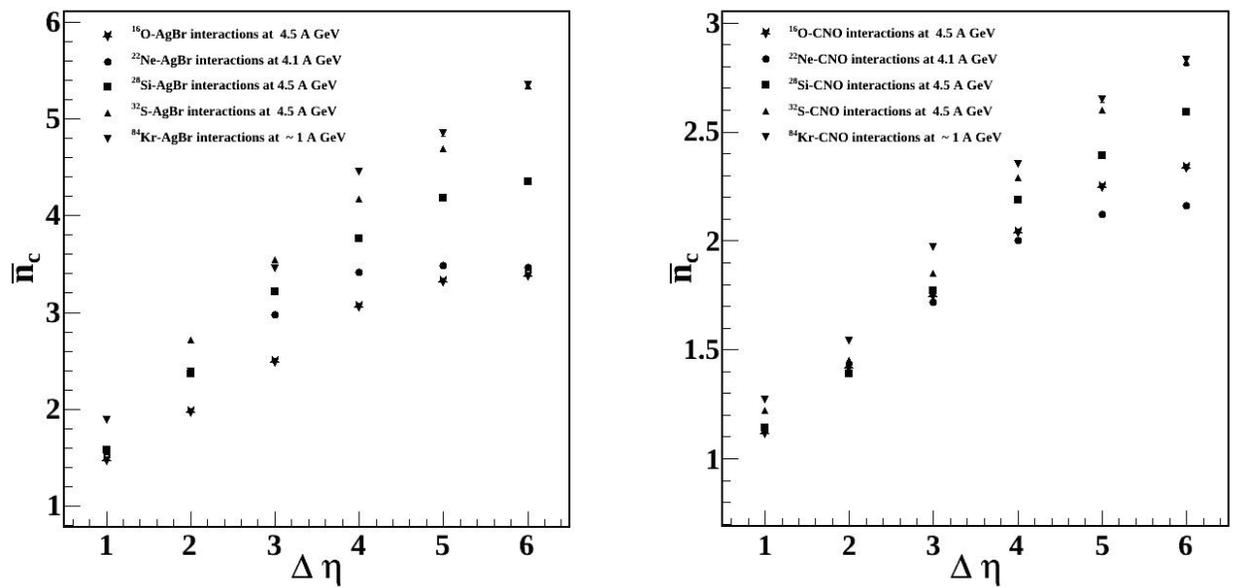

**Figure 3:** The variation of average number of particles per clan ($\bar{n}_c$) with respect to pseudo-rapidity interval ($\Delta\eta$) for different projectile with Ag Br target (left side) and CNO target (right side).